\documentclass[conference]{IEEEtran}
\usepackage{cite}
\usepackage{amsmath,amssymb,amsfonts,amsthm}
\usepackage{comment}

\usepackage{algorithmic}
\usepackage{algorithm}
\usepackage{enumerate}  
\DeclareMathOperator{\Tr}{Tr}

\usepackage{geometry}\usepackage[english]{babel}

\newtheorem{lemma}{\quad \textit{Lemma}}

\usepackage{subcaption}
\usepackage{graphicx}
\usepackage{textcomp}
\usepackage{optidef}
\usepackage{xcolor}

\begin{document}

\title{E-Healthcare Systems: Integrated Sensing, Computing, and Semantic Communication with Physical Layer Security}


\author{
\IEEEauthorblockN{Yinchao Yang\IEEEauthorrefmark{1},
Zhaohui Yang\IEEEauthorrefmark{2}\IEEEauthorrefmark{3},
Weijie Yuan \IEEEauthorrefmark{4},
Fan Liu \IEEEauthorrefmark{5},
Xiaowen Cao \IEEEauthorrefmark{6}, 
Chongwen Huang\IEEEauthorrefmark{2}\IEEEauthorrefmark{3},\\
Zhaoyang Zhang\IEEEauthorrefmark{2}\IEEEauthorrefmark{3},
and Mohammad Shikh-Bahaei\IEEEauthorrefmark{1}}

\IEEEauthorblockA{
$\IEEEauthorrefmark{1}$Department of Engineering, King's College London, London, UK \\
$\IEEEauthorrefmark{2}$College of Information Science and 
Electronic Engineering, Zhejiang University, Hangzhou, China\\
$\IEEEauthorrefmark{3}$Zhejiang Provincial Key Laboratory of Info. Proc., Commun. \& Netw. (IPCAN), Hangzhou, China\\
$\IEEEauthorrefmark{4}$ Department of Electrical and Electronic Engineering, \\Southern University of Science and Technology, Shenzhen 518055, China\\
$\IEEEauthorrefmark{5}$School of System Design
and Intelligent Manufacturing (SDIM), \\Southern University of Science and Technology, Shenzhen 518055, China\\
$\IEEEauthorrefmark{6}$School of Electronics and Information
Engineering, Shenzhen University, Shenzhen 518060, China\\
E-mails: 
yinchao.yang@kcl.ac.uk,
yang\_zhaohui@zju.edu.cn,
yuanwj@sustech.edu.cn,
liuf6@sustech.edu.cn, \\
caoxwen@szu.edu.cn,
chongwenhuang@zju.edu.cn,
ning\_ming@zju.edu.cn,
m.sbahaei@kcl.ac.uk
		}}

\maketitle

\begin{abstract} 
This paper introduces an integrated sensing, computing, and semantic communication (ISCSC) framework tailored for smart healthcare systems. The framework is evaluated in the context of smart healthcare, optimising the transmit beamforming matrix and semantic extraction ratio for improved data rates, sensing accuracy, and general data protection regulation (GDPR) compliance, while considering IoRT device computing capabilities. Semantic metrics such as semantic transmission rate and semantic secrecy rate are derived to evaluate data rate performance and GDPR risk, respectively, while the Cramér-Rao Bound (CRB) assesses sensing performance. Simulation results demonstrate the framework's effectiveness in ensuring reliable sensing, high data rates, and secure communication.
\end{abstract}


\IEEEpeerreviewmaketitle

\section{Introduction}
In recent times, the Internet of Things (IoT) has experienced a rapid and substantial evolution, drawing considerable interest. This transformative technology links everyday objects to the internet, enabling them to gather, share, and analyse data \cite{xing2020reliability}. With sensors, actuators, and communication capabilities, these interlinked devices can communicate with one another and with users, providing invaluable insights, optimising workflows, and boosting efficiency across various fields.

Meanwhile, robots are increasingly employed across diverse sectors and applications, ranging from manufacturing \cite{khan2023collaborative} and healthcare \cite{balasundaram2023internet} to education \cite{onesixg}. These robotic systems operate with objectives that complement those of the IoT, facilitating interaction, task execution, and autonomous behaviours. This synergy between robotic systems and the IoT, namely, the Internet of Robotic Things (IoRT), paves the way for enhanced interconnectivity and productivity in a wide array of fields. In the IoRT framework, robots are equipped with IoT devices like sensors and communication modules, enabling interaction with the environment and data exchange with other connected devices. This interconnected IoRT network can amass and analyse extensive data volumes, enabling real-time decision-making and adaptive behaviour across domains such as manufacturing, healthcare, agriculture, transportation, and smart cities \cite{ray2016internet,bobarshad2009m, shikh2007joint, nehra2010cross, ho2008design, shadmand2010multi, nehra2010spectral, kobravi2007cross, fang2021secure, olfat2008optimum, towhidlou2017improved, jia2020channel}. Nevertheless, the IoRT will substantially elevate the demands for environmental sensing and information transmission, potentially exacerbating spectrum shortages. The integration of sensing and communication (ISAC) emerges as a crucial enabler. Initially, early studies in this area predominantly focused on facilitating spectrum sharing between communication and sensing domains, such as communication and radar spectrum sharing. Subsequently, a more proactive approach has been advocated, promoting the joint utilisation of spectrum and hardware infrastructure between communication and sensing \cite{zhou2024near, zhang2019neural,olfat2008optimum, bobarshad2010low, ding2023distributed, ding2024joint}.

Despite the application of ISAC techniques, achieving high data rates for IoRT applications remains challenging. Semantic communication has recently emerged as a key approach to meet this demand, aided by advancements in artificial intelligence. Unlike traditional communication, which is bound by Shannon's capacity limit, semantic communication represents a paradigm shift that surpasses these constraints \cite{yang2023energy, xu2023edge, yang2024secure}. Semantic communication emphasises the extraction and transmission of message meaning rather than transmitting the entire message. 

Previous research efforts have primarily focused on either ISAC systems or semantic communications independently, without exploring the potential synergies between the two approaches. In the domain of ISAC, several works have been conducted. Yuan et al. \cite{yuan2020bayesian} explored ISAC beamforming design in vehicular networks, aiming to accurately track vehicles while maintaining conventional communication with them. Furthermore, Liu et al. \cite{liu2022learning} utilised deep learning techniques to find optimal ISAC beamforming in vehicular networks, aligning with the aforementioned design goals. On the other hand, in the field of semantic communications, Xie et al. \cite{xie2021deep} proposed a deep learning-based autoencoder system for the semantic communication of text messages. Building upon their work, Zhang et al. \cite{zhang2022semantic} introduced a semantic communication framework for wireless extended reality (XR) applications. Weng et al. \cite{weng2021semantic} focused on a deep learning-enabled semantic communication system for speech signals, while Wang et al. \cite{wang2023semantic} explored semantic communication for transmitting sensing results in the context of the Metaverse.


To address this research gap, this study introduces a novel framework that integrates semantic communication and ISAC functionalities, taking into account the computational capacity required for processing semantic messages. We designate this framework as \textbf{Integrated Sensing, Computing, and Semantic Communication} (ISCSC). By merging these complementary approaches, the framework facilitates the computation and transmission of semantic messages while capitalising on the inherent sensing capabilities of ISAC systems. The proposed framework holds applicability across a broad spectrum of IoRT applications. A particularly compelling use case lies within IoRT-enabled healthcare. In this scenario, robotic devices equipped with sensing and communication capabilities engage with IoT networks to monitor the environment and exchange semantic messages pertaining to patients in real-time. In this context, the knowledge base (KB) would encompass pertinent patient information and medical records, enabling precise interpretation and transmission of diagnostic data and treatment recommendations. In summary, our contributions are:
\begin{enumerate}
    \item We introduce the ISCSC framework, which is tailored for IoRT devices, streamlining the extraction, computation, and transmission of semantic information while capitalising on the inherent sensing functionalities of ISAC systems.

    \item We focus on one particular IoRT use case: Smart healthcare. In this scenario, we jointly design the transmit beamforming vectors and semantic extraction ratios while taking into account the computing capabilities of the IoRT device involved. The design aims are to maximise the data transmission rate and the sensing performances. In addition to that, we also consider secure transmission to prevent GDPR breaches.
\end{enumerate}

\section{System Model}

\begin{figure}[!t]
\centering
    \includegraphics[width=70mm]{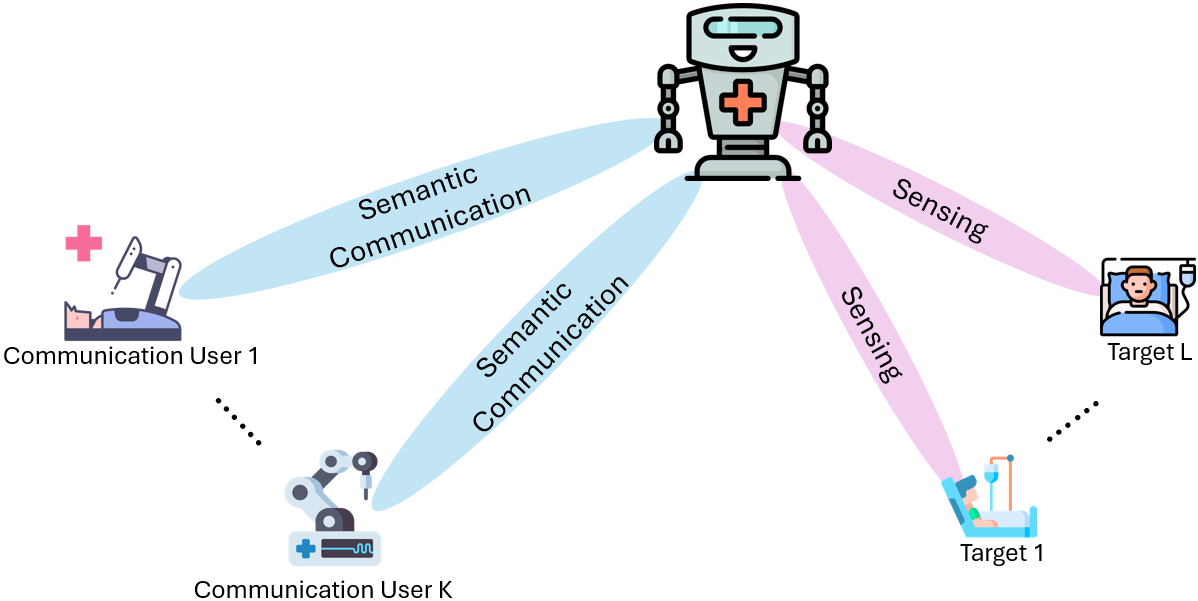}
    \caption{The system model of the considered use case}
    \label{smart healthcare}
\end{figure}

As shown in Fig. \ref{smart healthcare}, we consider an ISCSC system with a transmitter equipped with a uniform linear array (ULA) of $N$ antennas. The antennas are co-located for target detection and downlink semantic communication. The transmitter communicates with $K$ CUs, and each CU $k \in \mathcal{K}$ is equipped with a single antenna. Meanwhile, the transmitter aims to identify $L$ number of point-like targets, which is denoted by the set $\mathcal L$. During simultaneous communication and sensing activities, in compliance with GDPR, it is imperative to ensure that the messages transmitted to the CUs remain secure and are not disclosed to any unintended target. 

In the considered model, the received signal at the $k$-th CU can be characterised as follows:
\begin{equation}\label{eq1}
y_k = \mathbf{h}_k^H \mathbf{x} + n_k,
\end{equation}
where $\mathbf{h}_k\in\mathbb{C}^{N \times 1}$ is the channel vector for the $k$-th CU, $\mathbf{x}\in\mathbb{C}^{N \times 1}$ is the transmitted signal, and $n_k \sim \mathcal{CN}(0,\sigma^2_c)$ is the communication noise for the $k$-th CU. On the target side, the received signal can be formulated by
\begin{equation}\label{eq2}
y_l = \alpha_l \mathbf{a}^H(\theta_l) \mathbf{x} + n_l = \mathbf{h}_l^H \mathbf{x} + n_l,
\end{equation}
where $\alpha_l$ is the path-loss coefficient for the $l$-th target, $n_l \sim \mathcal{CN}(0,\sigma^2_r)$ is the sensing noise for the $l$-th target. The steering vector is denoted by $\mathbf{a}(\theta_l) \in \mathbb{C}^N$ with $\theta_l$ being the AOD of the $l$-th target, $|\theta_l| \leq\frac{\pi}{2}, \forall l$. The steering vector is formulated by 
\begin{equation}\label{eq3}
    \mathbf{a}^H(\theta_l) =  
    \begin{bmatrix}
    1 & e^{j2\pi \frac{d}{\lambda} \sin(\theta_l)} & \cdots & e^{j2\pi (N-1) \frac{d}{\lambda} \sin(\theta_l)}
    \end{bmatrix},
\end{equation}
where $\lambda$ is the wavelength and $d$ is the distance between any two adjacent antennas. 

The echo signal received by the transmitter after applying a matched filter can be formulated by
\begin{equation}\label{eq4}
    \mathbf{\hat{y}}_l = \beta_l \mathbf{a}(\theta_l) \mathbf{a}^H(\theta_l) \mathbf{x} + \mathbf{n}_l,
\end{equation}
where $\beta_l$ is the round-trip path loss coefficient and $\mathbf{n}_l \in \mathbb{C}^{N}$ is noise vector with zero mean and variance of $\sigma^2_e \mathbf{I}$. 

In the considered ISCSC system, both semantic communication signals and sensing signals are simultaneously transmitted. Thus, the transmitted signal $\mathbf{x}$ at the transmitter can be given by
\begin{equation}\label{eq5}
    \mathbf{x} = \mathbf{W} \mathbf{c} + \mathbf{R} \mathbf{z},
\end{equation}
where $\mathbf{W} \in \mathbb{C}^{N \times K}$ denotes the precoding matrix and $\mathbf{c} \in \mathbb{C}^{K}$ denotes the semantic message. Moreover, in \eqref{eq5}, $\mathbf{R} \in \mathbb{C}^{N \times L}$ represents the radar beamforming matrix and $\mathbf{z} \in \mathbb{C}^{L}$ is the sensing signal. As a result, the covariance matrix of the transmit waveform can be derived as follows:
\begin{equation}\label{eq9}
    \mathbf{R}_x = \mathbb{E}[\mathbf{x}\mathbf{x}^H] = \sum_{k=1}^K \mathbf{W}_k + \sum_{l=1}^L \mathbf{R}_l,
\end{equation}
where $\mathbf{W}_k = \mathbf{w}_k \mathbf{w}_k^H$ and $\mathbf{R}_l = \mathbf{r}_l \mathbf{r}_l^H$. 

\section{Performance Indicators}
\subsection{Semantic Communication}
In our previous work \cite{yang2024secure}, we proposed an explicit performance indicator for semantic communication by incorporating the  Bilingual Evaluation Understudy (BLEU) score. To provide a comprehensive understanding, we will briefly recap some important aspects of this work.

We establish the semantic transmission rate as the quantity of bits received by the CU following the extraction of semantic information, thus, the formulation is expressed as:
\begin{equation}\label{eq11}
    S_k = \frac{\iota}{\rho_k} \log_2(1+\gamma_k),
\end{equation}
where the parameter $\rho_k=\frac{\text{len}(c_k)}{\text{len}(m_k)}, 0\leq \rho_k \leq 1$ represents the semantic extraction ratio, and $\text{len}(\cdot)$ denotes the length of the message in words. Here, $\iota$ is a scalar value converting the word-to-bit ratio. The term $\gamma_k$ denotes the Signal-to-Interference-plus-Noise Ratio (SINR) for the $k$-th CU, expressed as:
\begin{equation}\label{eq12}
\footnotesize
\gamma_k= \frac{\Tr(\mathbf{h}_k \mathbf{h}_k^H \mathbf{W}_k)}{\Tr(\mathbf{h}_k \mathbf{h}_k^H \sum^K_{k'=1, k' \neq k}\mathbf{W}_{k'}) +  \Tr(\mathbf{h}_k \mathbf{h}_k^H \sum_{l=1}^L\mathbf{R}_l) + \sigma^2_c}.
\end{equation}

\begin{lemma}[\cite{yang2024secure}]
By choosing an appropriate global lower bound value for all $\text{BLEU}_k$ and appropriate values of $p_{g,k}$ for each CU, we can obtain the lower bound of $\rho_k$ as:\begin{equation}\label{eq14}
    \rho_k \geq \frac{1}{1 - \ln Q + \sum_{g=1}^G w_{g,k} \log p_{g,k}},
\end{equation}
where $w_{g,k}$ defines the weight of the g-grams for user $k$ and $p_{g,k}$ is the g-grams precision score for user $k$. The global lower bound of the BLEU score is denoted by $Q$.
\end{lemma}

By applying \textbf{\textit{Lemma 1}}, the lower bound of $\rho_k$ can be obtained. To prevent data breaches, we evaluate the worst-case Semantic Secrecy Rate (SSR) to assess the security level. Higher SSR means a lower chance of data breach events. Before formulating this, it's necessary to define the SINR of the $l$-th target (eavesdropper) in relation to the $k$-th CU:
\begin{equation}\label{eq15}
\footnotesize
    \Gamma_{l | k}  = \frac{\Tr(\mathbf{h}_l \mathbf{h}_l^H \mathbf{W}_k) }{\Tr(\mathbf{h}_l \mathbf{h}_l^H \sum_{k'=1, k'\neq k}^K \mathbf{W}_{k'}) + \Tr(\mathbf{h}_l \mathbf{h}_l^H \sum_{l=1}^L \mathbf{R}_l)  +  \sigma^2_r}.
\end{equation}

In the worst-case scenario, where the eavesdropper has an extensive KB similar to that of the transmitter and the legitimate CU, the semantic transmission rate for the $l$-th eavesdropper related to the $k$-th CU is determined as follows:
\begin{equation}\label{eq16}
    S_{l | k} = \frac{\iota}{\rho_k} \log_2 (1+\Gamma_{l | k}).
\end{equation}

In this way, the worst-case SSR of the $k$-th CU is formulated by
\begin{equation}\label{eq17}
    SSR_{k} = \min_{l \in L} [S_k - S_{l | k}]^+,
\end{equation}
where $[\cdot]^+$ means $\max(0, \cdot)$.

\subsection{Power Budget}
Deriving semantic information from a traditional message heavily relies on machine learning techniques. Hence, considering computational power as a component of the overall transmission power budget is crucial. In our previous work \cite{yang2024secure}, we proposed a natural logarithm function for computing the computational power:
\begin{equation}\label{eq18}
    P_{\text{comp}} =  \sum_{k=1}^K -F\ln(\rho_k),
\end{equation}
where F is a coefficient that converts a magnitude to its power. 

On the other hand, the communication and sensing energy consumption at the transmitter side is given by
\begin{equation}\label{eq19}
    P_{\text{c\&s}} = \Tr(\sum_{k=1}^K \mathbf{W}_k + \sum_{l=1}^L \mathbf{R}_l).
\end{equation}

Additionally, the overall transmission power consumption is limited to the power budget:
\begin{equation}\label{eq20}
    P_{\text{comp}} + P_{\text{c\&s}} \leq P_t,
\end{equation}
with $P_t$ being the total power budget.

\subsection{Cramér-Rao Bound}
Typically, mean square errors (MSE) between the estimated angle and the actual angle serve as metrics for evaluating sensing performance. However, deriving a closed-form expression for MSE can be challenging \cite{1703855}. As an alternative, for assessing static target sensing performance, we utilize the \textit{Cramér-Rao Bound} (CRB), which offers a lower bound of the MSE and can be expressed in a closed form. The formulation of CRB has been derived in \cite{1703855}; hence, we provide a brief description below. Defining the parameters to be estimated as $\xi = [\theta_l, \beta_l]$, the \textit{Fisher Information Matrix} (FIM) is given by:
\begin{equation}\label{eq21}
    \mathbf{J}_l = 
    \begin{bmatrix}
        {J}_{\theta_l \theta_l} & \mathbf{J}_{\theta_l \beta_l}\\
        \mathbf{J}_{\theta_l \beta_l}^T & \mathbf{J}_{\beta_l \beta_l}
    \end{bmatrix}.
\end{equation}

For simplicity, let use define two variables $\mathbf{B}_l = \mathbf{a}(\theta_l) \mathbf{a}^H(\theta_l)$ and $\Dot{\mathbf{B}}_{\theta_l} = \frac{\partial \mathbf{B}_l}{\partial \theta_l}$. Hence, we can obtain the following equations:
\begin{equation}\label{eq22}
    {J}_{\theta_l \theta_l} = \frac{2T|\beta_l|^2}{\sigma^2_r} \Tr(\dot{\mathbf{B}}_{\theta_l} \mathbf{R}_x \dot{\mathbf{B}}_{\theta_l}^H),
\end{equation}
\begin{equation}\label{eq23}
    \mathbf{J}_{\theta_l \beta_l} = \frac{2T\beta_l^*}{\sigma^2_r} \Re\{\Tr(\mathbf{B}_l \mathbf{R}_x \dot{\mathbf{B}}^H_{\theta_l})\} [1,j],
\end{equation}
\begin{equation}\label{eq24}
    \mathbf{J}_{\beta_l \beta_l} = \frac{2T}{\sigma^2_r} \Tr(\mathbf{B}_l \mathbf{R}_x \mathbf{B}_l^H) \mathbf{I}.
\end{equation}

Therefore, the CRB of $\theta_l$ is formulated by
\begin{equation}\label{eq25}
    CRB(\theta_l) = {\mathbf{J}_l^{-1}}_{[1,1]}=({J}_{\theta_l \theta_l} - \mathbf{J}_{\theta_l \beta_l} \mathbf{J}^{-1}_{\beta_l \beta_l} \mathbf{J}_{\theta_l \beta_l}^T)^{-1},
\end{equation}
where $\mathbf{A}_{[1,1]}$ means the element of matrix $\mathbf{A}$ in row 1 and column 1. 

\section{Integrated Sensing, Computing and Semantic Communication Design for Smart Healthcare}

In this section, we assume the CSI of all CUs are perfect as they are non-portable medical equipment. In addition, we assume the locations of all patients (eavesdroppers) are available at the transmitter but may contain errors. 

\subsection{Problem Formulation}

In our approach to jointly designing beamformers and semantic extraction ratios, the objective is to maximize the sum worst-case SSR while simultaneously minimizing the sum CRB of eavesdropper angles. This dual optimization aims to enhance semantic security while enabling more precise estimations of eavesdropper locations. Such improvements bolster overall system security.


Therefore, the optimization problem is given by
\begin{subequations}\label{eq26}
\begin{align}
    \max_{\mathbf{W}_k, \mathbf{R}_l, \rho_k}\; &  \kappa_1 \sum_{k=1}^K SSR_k - \kappa_2 \sum_{l=1}^L CRB(\theta_l)\label{eq26a}\\
    \text{s.t.} \quad & \frac{1}{1 - \ln Q + \sum_{g=1}^G w_{g,k} \log p_{g,k}} \leq \rho_k \leq 1, \forall k,\label{eq26b}\\
    & S_k \geq \varsigma, \forall k,\label{eq26c}\\
    &  P_{\text{comp}} + P_{\text{c\&s}} \leq P_t,\label{eq26d}\\
    &\mathbf{W}_k \succeq 0, \mathbf{W}_k = \mathbf{W}_k^H,  \forall k,\label{eqGOPe}\\
    & \mathbf{R}_l \succeq 0, \mathbf{R}_l = \mathbf{R}_l^H, \forall l,\label{eqGOPf}\\
    & \text{rank}(\mathbf{W}_k) = 1,  \forall k,
\end{align}
\end{subequations}
where $\kappa_1$ and $\kappa_2$ are the trade-off terms. Since the user CSIs are perfect, and the target channels are imperfect, the worst-case SSR in \eqref{eq26a} can be modified to:
\begin{equation}\label{eq27}
    SSR_{k} =  [S_k - \max_{l\in L} S_{l | k}]^+.
\end{equation}

\subsection{Algorithm Design}

By introducing auxiliary variables $\lambda_k, \forall k$, we can transform the first part of \eqref{eq26a} into the following form:
\begin{equation}\label{eq28}
    \max_{\mathbf{W}_k, \mathbf{R}_l, \rho_k, \lambda_{k}} \kappa_1 \sum_{k=1}^K  (S_k - \frac{\iota}{\rho_k} \log_2(1+\lambda_{k})),
\end{equation}
with an additional constraint of
\begin{equation}\label{eq29}
    \Gamma_{l | k} \leq \lambda_{k}, \forall k, \forall l.
\end{equation}

However, \eqref{eq28} is a non-concave function due to the logarithm terms in the equation. We apply first-order Taylor expansion which leads to the following results:
\begin{equation}\label{eq30}
S_k \overset{\Delta}{=} \frac{\iota}{\rho_k} \Big( \log_2(A_k) - \log_2(B_k^i) - \frac{1}{B_k^i \ln(2)} (B_k - B_k^i) \Big),
\end{equation}
and
\begin{equation}\label{eq31}
    \begin{aligned}
        &-\frac{\iota}{\rho_k}\Big( \log_2(1 + \lambda_{k}) \Big)\\
        &\overset{\Delta}{=} -\frac{\iota}{\rho_k} \Big( \log_2(C_{k}^i) + \frac{1}{C_{l | k}^i \ln(2)} (C_{k} - C_{k}^i) \Big),
    \end{aligned}
\end{equation}
where superscript $i$ means the value of a variable in the $i$-th iteration. Additional parameters are listed below:
\begin{equation}
    \begin{cases}
        A_k = \mathbf{h}_k^H \sum_{k=1}^K \mathbf{W}_k \mathbf{h}_k + \mathbf{h}_k^H \sum_{l=1}^L \mathbf{R}_l \mathbf{h}_k + \sigma_c^2, \\
        B_k = \mathbf{h}_k^H \sum_{k'=1, k'\neq k}^K \mathbf{W}_{k'} \mathbf{h}_k + \mathbf{h}_k^H \sum_{l=1}^L \mathbf{R}_l \mathbf{h}_k + \sigma_c^2, \\
        C_{k} = 1 + \lambda_{k}.
    \end{cases}
\end{equation}

Since the estimations of the eavesdropper's location contain errors, the eavesdropper's channels ($\mathbf{h}_l, \forall l$) have errors. We assume the channel errors are bounded by a spherical region:
\begin{equation}\label{eq32}
    \mathcal{H}_l := \{ (\mathbf{\hat{h}}_l + \mathbf{u}_l)^H \; | \; ||\mathbf{u}_l|| \leq \varepsilon_l\}, \forall l,
\end{equation}
where $\varepsilon_l \geq 0$ corresponds to the radius of $\mathcal{H}_l$, and $\mathbf{h}_l$ is the actual channel and $\hat{\mathbf{h}}_l$ represents the estimated channel, while $\mathbf{u}_l$ represents the error vector. Therefore, the SNR for the $l$-th eavesdroppers becomes
\begin{equation}\label{eq33}
    \Gamma_{l | k} = \frac{(\hat{\mathbf{h}}_l + \mathbf{u}_l)^H \mathbf{W}_k (\hat{\mathbf{h}}_l + \mathbf{u}_l)}{(\hat{\mathbf{h}}_l + \mathbf{u}_l)^H \sum_{l=1}^L \mathbf{R}_l (\hat{\mathbf{h}}_l + \mathbf{u}_l) + \sigma_r^2 }.
\end{equation}

By using \eqref{eq33}, we can apply the S-procedure on the constraint \eqref{eq29}. Through denoting $(\mathbf{W}_k - \lambda_{k}\sum_{l=1}^L \mathbf{R}_l)$ as $\mathbf{E}_{k}$, \eqref{eq29} can be replaced by:
\begin{equation}\label{eq34}
\begin{bmatrix}
    t_{k} \mathbf{I} - \mathbf{E}_k & - \hat{\mathbf{h}}_l^H \mathbf{E}_{k}\\
    -\mathbf{E}_{k} \hat{\mathbf{h}}_l & -t_{k}\varepsilon_l^2 - \hat{\mathbf{h}}_l^H \mathbf{E}_{k} \hat{\mathbf{h}}_l + \lambda_{k} \sigma_r^2
\end{bmatrix}
\succeq 0 
\end{equation}
where $t_k \geq 0$ is an auxiliary variable relying on the S-procedure. 

Lastly, to tackle the non-concave function CRB, we transform it as a constraint of
\begin{equation}\label{eq35}
    \begin{bmatrix}
        {J}_{\theta_l \theta_l}-{U}_{l} & \mathbf{J}_{\theta_l \beta_l}\\
        \mathbf{J}_{\theta_l \beta_l}^T & \mathbf{J}_{\beta_l \beta_l} 
    \end{bmatrix} \succeq 0, \forall l,\\
\end{equation}
with the CRB part in \eqref{eq26a} becomes $-\kappa_2 (\sum_{l=1}^L U_l^{-1})$ and $U_l \geq 0$ is a new variable introduced.

Hence, by dropping the rank-one constraint using Semi-definite relaxation (SDR), \eqref{eq26} is transformed to:
\begin{subequations}\label{eq36}
\begin{align}
    \max_{\mathbf{\Psi}} \; &  \kappa_1 \sum_{k=1}^K  \frac{\iota}{\rho_k} \Big( \log_2(A_k) - \log_2(B_k^i) \nonumber\\
    &- \frac{1}{B_k^i \ln(2)} (B_k - B_k^i) -  \log_2(C_{k}^i) \nonumber\\
    &- \frac{1}{C_{k}^i \ln(2)} (C_{k} - C_{k}^i)\Big) - \kappa_2 (\sum_{l=1}^L U_l^{-1}) \label{eq36a}\\
    \text{s.t.} \quad & \frac{1}{1 - \ln Q + \sum_{g=1}^G w_{g,k} \log p_{g,k}} \leq \rho_k \leq 1, \forall k,\label{eq36b}\\
    & \frac{\iota}{\rho_k} \Big( \log_2(A_k) - \log_2(B_k^i) - \frac{1}{B_k^i \ln(2)} (B_k \nonumber \\&- B_k^i) \Big)\geq \varsigma, \forall k, \label{eq36c}\\
    & \begin{bmatrix}
    t_{k} \mathbf{I} - \mathbf{E}_{k} & - \hat{\mathbf{h}}_l^H \mathbf{E}_{k}\\
    -\mathbf{E}_{k} \hat{\mathbf{h}}_l & -t_{k} \varepsilon_l^2 - \hat{\mathbf{h}}_l^H \mathbf{E}_{k} \hat{\mathbf{h}}_l + \lambda_{k} \sigma_r^2
    \end{bmatrix}
    \succeq 0, \forall k, \forall l,\label{eq36d} \\
    & \begin{bmatrix}
        {J}_{\theta_l \theta_l}-U_{l} & \mathbf{J}_{\theta_l \beta_l}\\
        \mathbf{J}_{\beta_l \theta_l} & \mathbf{J}_{\beta_l \beta_l} 
    \end{bmatrix} \succeq 0, \forall l,\label{eq36e}\\
    & P_{\text{comp}} + P_{\text{c\&s}} \leq P_t, \label{eq36f}\\
    &\eqref{eqGOPe}, \eqref{eqGOPf}, \label{eq36g}
\end{align}
\end{subequations}
where $\mathbf{\Psi} = [\mathbf{W}_k, \mathbf{R}_l, U_l, \rho_k, \lambda_k, t_k]$. To solve \eqref{eq36}, we consider using alternating optimization:

\subsubsection*{\textbf{Step 1}} With given $\rho_k$ and $ \lambda_k$ in problem \eqref{eq36}, the joint sensing and communication beamforming optimization problem can be given by
\begin{subequations}\label{eq37}
\begin{align}
    \max_{\mathbf{W}_k, \mathbf{R}_l, U_l, t_k}\; &  \kappa_1 \sum_{k=1}^K \frac{\iota}{\rho_k} \Big( \log_2(A_k) - \log_2(B_k^i) - \frac{1}{B_k^i \ln(2)} \nonumber\\ &(B_k - B_k^i) - \log_2(1+\lambda_{k}) \Big) -\kappa_2 (\sum_{l=1}^L U_l^{-1}) \label{eq37a}\\
    \text{s.t.} \quad & \eqref{eq36c}-\eqref{eq36g},
\end{align}
\end{subequations}
where $B_k^i$ is updated iteratively. Optimization problem \eqref{eq37} is convex and can be effectively solved via the standard convex optimization tool.

\subsubsection*{\textbf{Step 2}} With fixed $\rho_k$ as well as the obtained values of $\mathbf{W}_k$ and $\mathbf{R}_l$ from Step 1, the optimization of $\lambda_{k}$ is 
\begin{subequations}\label{eq38}
\begin{align}
    \max_{\lambda_k}\; &  \sum_{k=1}^K \frac{\iota}{\rho_k} \Big( \log_2(A_k) - \log_2(B_k) \nonumber\\ 
    &-\log_2(C_{k}^i) 
    -\frac{1}{C_{k}^i \ln(2)} (C_{k} - C_{k}^i) \Big) \label{eq38a}\\
    \text{s.t.} \quad & \Gamma_{l | k} \leq \lambda_k, \forall k, \forall l, 
\end{align}
\end{subequations}
where $A_k$, $B_k$ and $\Gamma_{l | k}$ are known values. $C_k^i$ is updated iteratively and $C_k = 1 + \lambda_k$ is the variable. Problem \eqref{eq38} is convex and can be effectively solved. 

\subsubsection*{\textbf{Step 3}} With obtained values of $\mathbf{W}_k$, $\mathbf{R}_k$, $\lambda_k$, the optimization of semantic extraction ratio $\rho_k$ is
\begin{subequations}\label{eq39}
\begin{align}
    \max_{\rho_k} &  \sum_{k=1}^K  \frac{\iota}{\rho_k} \Big( \log_2(A_k) - \log_2(B_k) - \log_2(C_{k}) \Big) \label{eq39a}\\
    \text{s.t.} \quad & \eqref{eq36b}, \eqref{eq36f}, \\
    & \frac{\iota}{\rho_k} \Big( \log_2(A_k) - \log_2(B_k)\Big)\geq \varsigma, \forall k, \label{eq39v}
\end{align}
\end{subequations}
where $A_k$, $B_k$ and $C_k$ are known values. Problem \eqref{eq39} is also convex and can be solved through the bisection method.

\subsubsection*{\textbf{Step 4}}
As a last step, we use Gaussian randomisation to recover rank-one solutions. The detailed procedures for solving problem \eqref{eq36} are shown in Algorithm \ref{alg:1}.  

\subsection{Algorithm Complexity and Convergence}
The complexity of Algorithm 1 is $\mathcal{O}(I_1 I_2 (KN^2)^3)$, where $I_1$ is the number of outer loop iterations and $I_2$ is the number of iterations for solving \eqref{eq37}.

\begin{algorithm}
\caption{Iterative Sensing, Communication, and Semantic Optimization Algorithm}\label{alg:1}
\begin{algorithmic}[1]
\STATE Initialize $\rho_k^0$, $\mathbf{W}_k^0$, $\mathbf{R}_k^0$, $\lambda_k$, $B_k^0$ and $C_k^0$.
\REPEAT
\REPEAT
    \STATE Solve convex problem \eqref{eq37} with the existing cvx toolbox. 
    \STATE Update $\mathbf{W}_k^{i+1}$ and $\mathbf{R}_k^{i+1}$.
\UNTIL{$|\mathbf{W}_k^{i+1} - \mathbf{W}_k^i| \leq \varrho_1$, $|\mathbf{R}_k^{i+1} - \mathbf{R}_k^i| \leq \varrho_2$.}
\REPEAT
    \STATE Solve convex problem \eqref{eq38} with the existing cvx toolbox. 
    \STATE Update $\lambda_k^{i+1}$.
\UNTIL{ $|\lambda_k^{i+1} - \lambda_k^i| \leq \varrho_3$.}
    \STATE Apply the bisection method and update $\rho_k^{i+1}$.
\UNTIL{converge}
\STATE Apply Gaussian randomisation.
\end{algorithmic}
\end{algorithm}

The $n$-th iterate $\mathbf{W}^{i}, \mathbf{R}^{i}$ yield a non decreasing sequence of $A^{n} \geq ... \geq A^{0}$ and $B^{n} \geq ... \geq B^{0}$, $\lambda_k^n$ yield a non decreasing sequence of $C^{n} \geq ... \geq C^{0}$, and $\rho_k^{n} \geq ... \geq \rho_k^0$, therefore, Algorithm \ref{alg:1} must converge.

\section{Numerical Results}

In this section, we present numerical findings to evaluate the effectiveness of our proposed designs. Our experimental setup assumes the transmitter employs ULAs with half-wavelength spacing, featuring a total of 20 antennas. The targets are situated at angular coordinates of [$-35^\circ$, $5^\circ$, $40^\circ$], with CUs positioned at [$-30^\circ$, $20^\circ$]. Noise power is standardized to $-30$ dBm, indicated by $\sigma^2_r = \sigma^2_c = -30$ dBm. A total power budget of 20 dBm is allocated. In \eqref{eq11}, we set $\iota = 1.1$ and $\kappa = 0.5$ to define the trade-off term (sensing and communication are equally important). Additionally, we establish lower bounds of 0.4 for $\rho_1$ and 0.33 for $\rho_2$. We further assign values of $\varsigma = 1$ and $\varepsilon_l = 0.01$.

\begin{figure}[!t]
\centering
\includegraphics[width=60mm]{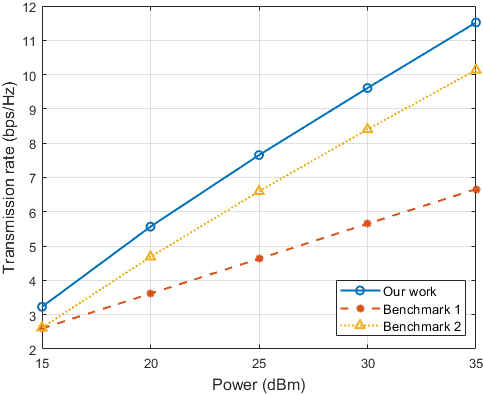}
\caption{Transmission rate}
\label{case 1 TR}
\end{figure}

\begin{figure}[!t]
\centering
\includegraphics[width=60mm]{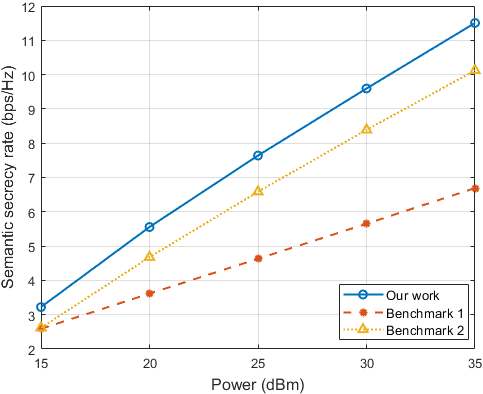}
\caption{Semantic secrecy rate}
\label{case 1 SSR}
\end{figure}

\begin{figure}[!t]
    \centering
    \includegraphics[width=60mm]{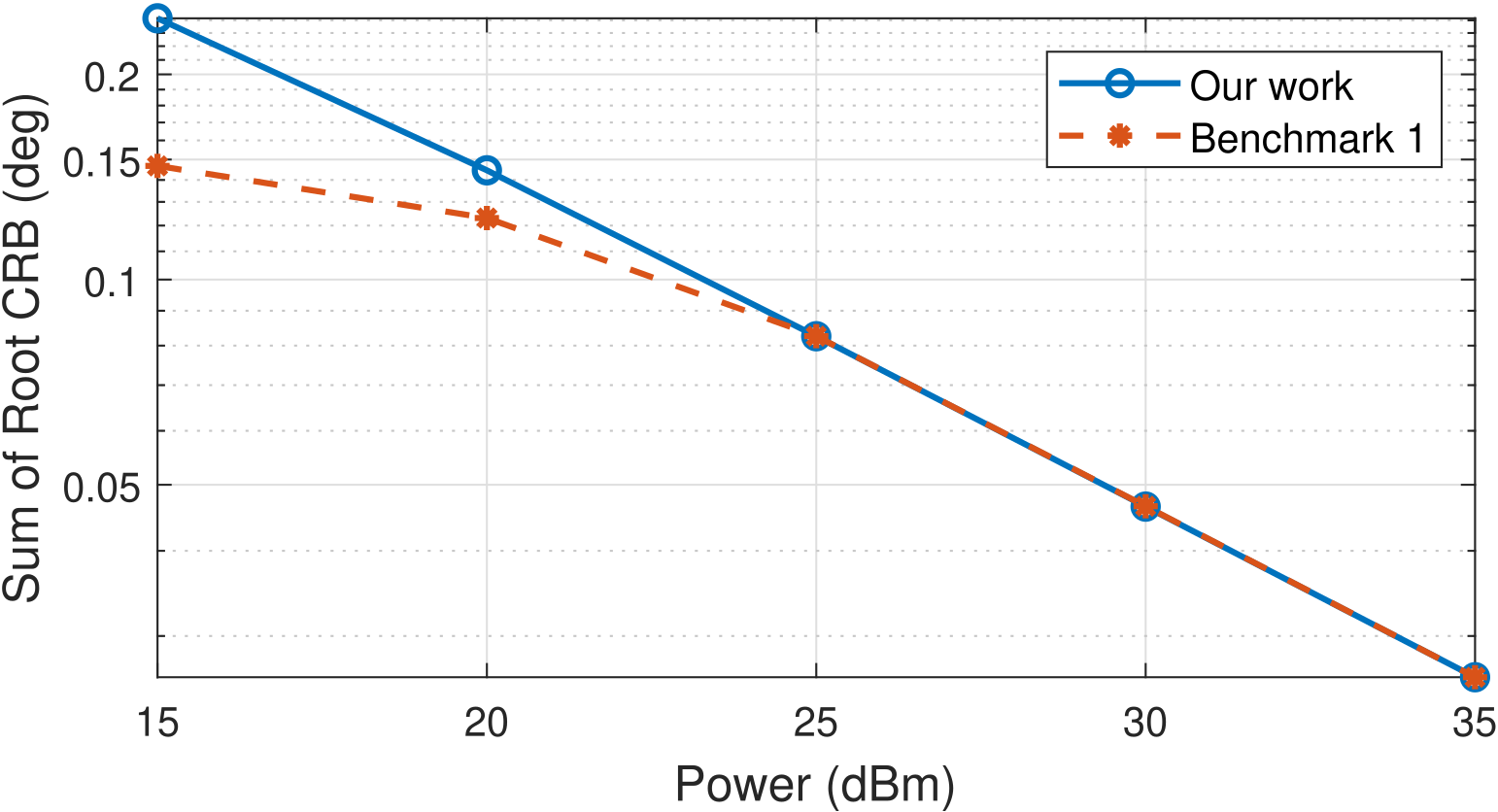}
    \caption{Sum of RCRB achieved in smart healthcare}
    \label{RCRB}
\end{figure}

In our evaluation, we first establish two benchmarks. Benchmark 1 is based on conventional ISAC designs, such as the approach proposed in \cite{jia2023physical}, applied to the IoRT-enabled smart healthcare model. Benchmark 2 represents our algorithm with the parameter $\rho_k$ fixed at 1. Fig. \ref{case 1 TR} and Fig. \ref{case 1 SSR} illustrate the performance of semantic communication and the semantic secrecy rate achieved. Compared to Benchmarks 1 and 2, the incorporation of semantic communication techniques leads to an improvement in the transmission rate by approximately $24\%$ when the power level is set to 15 dBm. This improvement becomes more pronounced as the power budget increases, as depicted in Fig. \ref{case 1 TR}. At a power budget of 35 dBm, our design enhances the transmission rate by around $70\%$ compared to Benchmark 1. When contrasted with Benchmark 2, the transmission rate can be improved by $14\%$ through the application of semantic communication techniques. The improvement in the transmission rate directly translates to an enhancement in the secrecy rate achieved by our design, as illustrated in Fig. \ref{case 1 SSR}.

The sensing performance achieved by our proposed approach and Benchmark 1 is illustrated in Fig. \ref{RCRB}, which depicts the sum of the RCRB values. When the power budget is set at 15 dBm, Benchmark 1 exhibits a slightly better sensing performance compared to our approach, as evidenced by a lower sum of RCRB values. However, as the power budget increases, the sensing performances of our approach and Benchmark 1 converge, with negligible differences observed. It is worth noting that a lower sum of RCRB values corresponds to better sensing performance, as the RCRB provides a lower bound on the variance of unbiased estimators, thereby serving as a metric for achievable estimation accuracy. Therefore, while Benchmark 1 holds a marginal advantage in sensing performance at lower power levels, our proposed approach matches its performance as the available power budget increases.

\section{Conclusion}
This paper explores the joint design of transmit beamforming vectors and semantic extraction ratios for an ISCSC-based smart healthcare system. The performance indicators are used to evaluate semantic communication performance and sensing performance. Tailored algorithms are developed to accommodate the channel errors to ensure the quality of communication. Simulation results demonstrate the proposed framework and algorithms intelligently balance sensing accuracy, communication throughput, and data privacy for smart healthcare applications.

\bibliographystyle{ieeetr}
\bibliography{bib}

\end{document}